# ROOF FALL HAZARD DUE TO BLASTING ACTIVITY IN THE LIGHT OF NUMERICAL MODELING AND UNDERGROUND MEASUREMENTS

**Witold Pytel**[1], **Krzysztof Fuławka**[1], **Piotr Mertuszka**[1], **Marcin Szumny**[1],

[1]KGHM CUPRUM Ltd. Research and Development Centre, **Poland**

**ABSTRACT**

*One of the major problems associated with exploitation of the copper ore deposit in underground mines in Poland is the local disturbance in a state of stable equilibrium manifested in a sudden release of strain energy stored in the deformed rock mass. It occurs mainly in the form of dynamic events which may result in rockbursts and roof falls. In order to face this threats, a number of organisational and technical prevention methods are applied in mines. It should be also noted that the greatest difficulties with the roof control are observed in the vicinity of active mining fronts, where the highest deformations are observed. The detonation of explosives generates a propagating shock wave which may cause a serious damage to a material body that is encountered on its way. Thus, a number of doubts during the mining operation emerged, that simultaneous firing of group of mining faces may have the negative impact on the condition of applied roof support and condition of roof strata as well The article discusses geomechanical influence of multi-faces blasting on immediate roof strata condition through the mutual comparison of the instrumented bolts monitoring data and the computer simulations results. The numerically assessed stress/strain field in the near vicinity of the blasting works operation has proved to be in close agreement with the field measured data. In the considered mining conditions both the numerical approach and field strain/stress monitoring indicated the low effect of production blasting on the immediate roof fall potential.*

1. ## INTRODUCTION

A major natural hazard associated with mining of the copper ore deposit located in the Legnica-Głogów Copper Belt area is the dynamic phenomena occurrences, physically manifested as seismic tremors [1]. Seismic activity increases with increasing of the mined out areas as well their depths and this may lead to higher number of rockbursts and roof falls occurrences [2]. In order to face this threats, a number of organisational and technical prevention methods are applied. Currently, the most effective approach is utilizing the active prevention methods based on blasting works which allow for reduction of the stress concentrations and cracks formation within the overloaded rock mass [3].

The most often practised rockburst prevention methods in Polish copper mines are based on blasting works and are applied as a group winning blasting, i.e. simultaneous firing of dozens of mining faces, in order to amplifying the elastic wave and, in result, to triggering of high-energy tremors [4, 5]. Based on the previous experiences it was noticed that a significant number of recorded dynamic events can be clearly and directly explained by the effects of those methods [6]. Therefore, it can be assumed that the greater number of simultaneously fired mining faces will contribute to improvement of effectiveness of high-energy tremors provocation.

However, some doubts as to the method's safety have emerged. Namely, there is a concern that the simultaneous detonation of higher amount of explosives will contribute to increasing the roof fall risk, mainly within the mine workings located in the close vicinity of exploitation front. It was expected that seismic wave generated by blasting may have the negative impact on the condition of applied roof support and can be cause of instabilities within immediate roof strata.

The authors' overall goal was to assess whether the multi-face production blasting which are aimed to increase the capability of inducing stress relief in the rock mass, may have negative effects on the roof fall hazard. In this paper, the influence of multi-faces blasting on roof strata condition were analyzed based on the results of computer simulations. The analyses have been supported by a 3D FEM modelling. Results of the numerical model solution were verified with respect to stress values measured within the immediate roof strata using the 3D instrumented rock bolts.

## 2. MATERIALS AND METHODS

To determine the impact of seismic wave generated by blasting works on stress state within the immediate roof strata, one of the archival blasting operation has been selected. It consisted of simultaneous firing of 15 faces with the total amount of used explosives of about 900 kg (60 kg per mining face). Since the highest amount of explosives fired with same delay is located in the cut holes, those blastholes were considered exclusively. It was therefore assumed, that the total analysis period of time was 0.5 s. Figure 1 shows the applied drilling and firing pattern. The cut holes are market as "1".

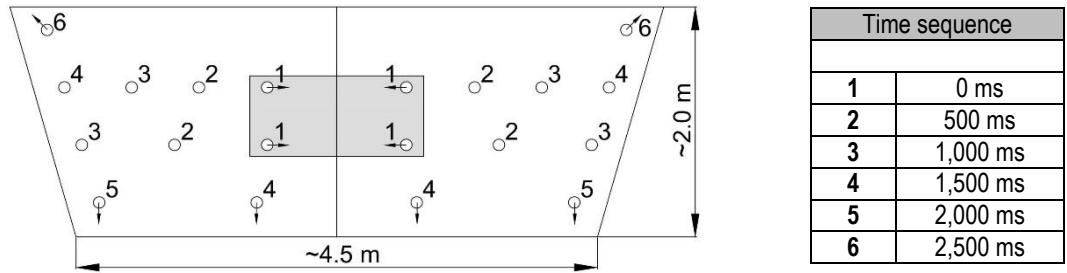

Fig. 1. V-cut pattern with mining face dimensions and delay sequence

The deposit within considered mining panel is located ca. 750 m beneath the surface and is classified as stratoidal and single-level. The roof strata consist of slightly weathered and cavernous, fine-stratum dark grey dolomites with grey clay. Lithostratigraphy within the considered mining panel is presented in Figure 2.

In order to determine stress values distribution within the immediate roof strata of considered mining panel, a set of two instrumented rock bolts were installed there. They were located in close proximity to active mining front. Therefore, the most suitable location for conducting such measurements was the gallery between two active mining panels (see Figure 3).

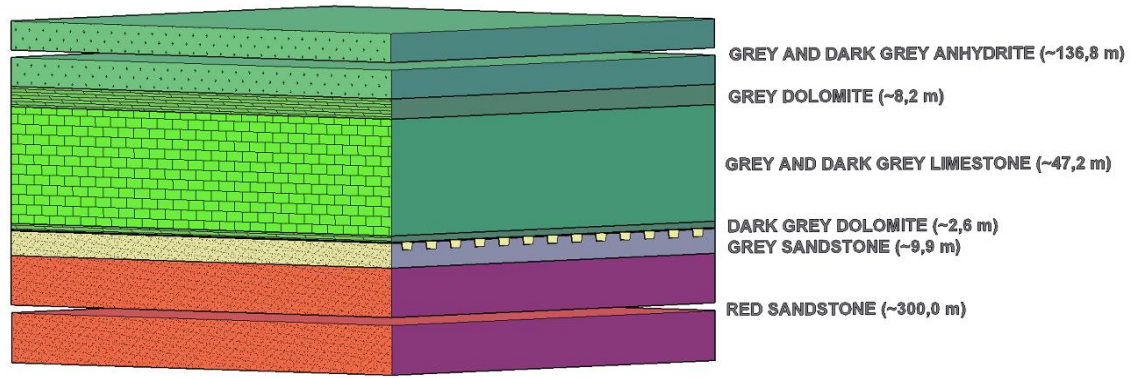

**Fig. 2. Simplified geological profile within analysed mining panel**

Two instrumented rock bolts were installed in accordance with the standard bolting pattern, i.e. spacing of 1.5 m x 1.5 m. It allowed to precisely reflect the impact of blasting works on the stability of roof layers. In order to obtain reliable data, monitoring system was installed in the forefield of the mining front line. The minimum distance between fired face and measuring station was 51 m, while maximum distance exceeded 400 m. Locations of the instrumented rock bolts installation is presented in Figure 3.

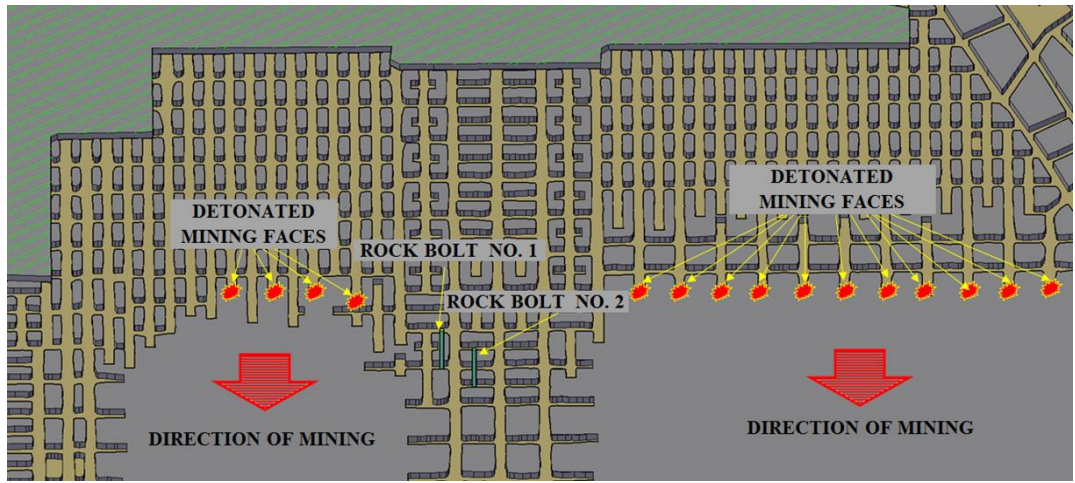

**Fig. 3. Location of monitoring stations and fired faces in relation to the actual mining geometry**

The instrumented rock bolts used during underground tests allowed for simultaneous stress data acquisition in four perpendicular directions at five measuring levels. High sensitivity of the applied strain gauges and sampling frequency of 10 Hz allowed for the full mapping of forces acting on the bolts' rod with very high accuracy. Scheme of used instrumented rock bolt and examples of recorded axial stresses, given separately for each measuring level, are presented in Figure 4.

The measured stresses were compared with results of numerical 3D FEM simulations. For this purpose NEi/NASTRAN computer program code utilizing FEM in three dimensions has been applied. It was assumed that all of the materials reveal linear elastic characteristics, except for rocks comprised within pillars which exhibit elastoplastic behaviour with strain softening. The boundary conditions of the entire numerical models were described by displacement-based relationships (no vertical displacements of the bottom layer in the model and no movements in the direction perpendicular to the lateral walls). The multi-plate overburden model has been assumed as a basic physical model, which consists of several homogeneous rock plates reflecting the real lithology within considered area. A room-and-pillar mining system with roof deflection was assumed.

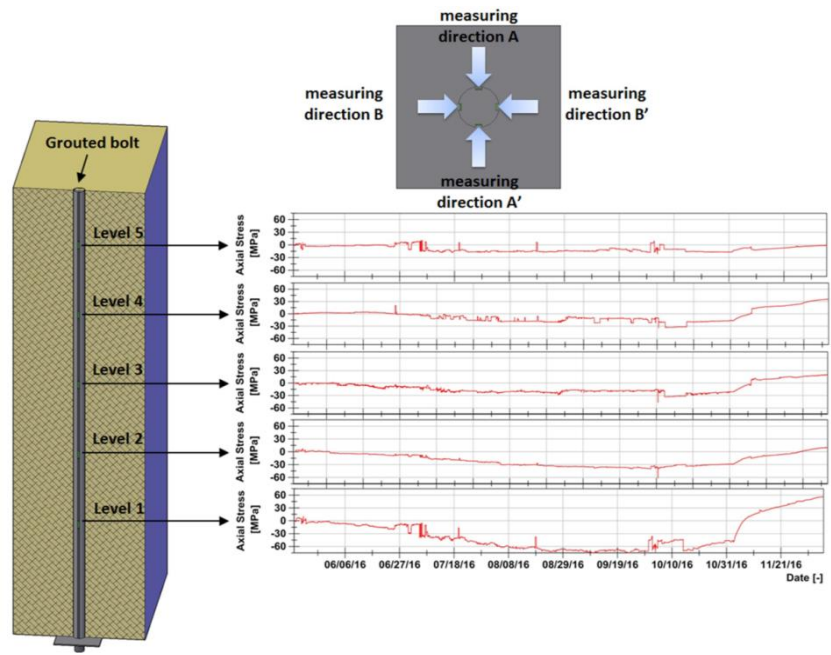

**Fig. 4. Instrumented rock bolt scheme with the data samples**

The detonation of explosives in the mining faces were simulated by applying hydrostatic pressure within a finite elements increasing from 0 to 100 MPa within 1 ms and decreasing after detonation to zero [7].

## 3. ANALYSIS OF RESULTS

For the purpose of analysis, the immediate roof strata which consists of dolomite was divided into eleven smaller layers. Six of them (no. 1÷6) were located within the rock bolt support level, i.e. up to 1.5 m above the roof. Thickness of the layers located at this level varied from 0.18 m to 0.32 m. The remaining five layers (no. 7÷11) with thickness between 0.80 and 2.30 m, were located above the support level (Tab. 1). The stratification of the dolomite stratum allowed to determine the changes of stresses along the rock bolt's rod. High accuracy of dynamic numerical analysis was achieved by relatively short time step ($\Delta t = 0.0005$ s). Since the instrumented rock bolts were used mainly to monitor the roof fall hazard, the sampling frequencies of applied measuring systems were 10 Hz which was the most suitable value from the power supply efficiency and amount of the data point of view. Consequently, signals recorded in situ and obtained numerically have different time steps. To match the resolutions of the FEM results and measured values, the Savitzky-Golay (S-G) filter was applied.

Tab. 1. Parameters of dolomite stratum used in numerical analysis

| Material $m_i$ | Layer ID | Layer thickness (m) | $C_o^{(n)} \div C_o^{(r)}$ (MPa) | $T_o^{(n)} \div T_o^{(r)}$ (MPa) | $E_s \div E^{(r)}$ (MPa) | Poisson's ratio $\nu$ | Hoek-Brown material constants | | |
|---|---|---|---|---|---|---|---|---|---|
| | | | | | | | $m_b$ | s | a |
| Dolomite stratum above the support level | 11 | 2.30 | 146.7 ÷ 61.4 | 9.9 ÷ 0.85 | 51400 ÷ 12850 | 0.24 | 4.90 | 0.33 | 0.5 |
| | 10 | 2.30 | | | | | | | |
| | 9 | 2.00 | | | | | | | |
| | 8 | 0.80 | | | | | | | |
| | 7 | 0.30 | | | | | | | |
| Dolomite stratum within the support level | 6 | 0.18 | | | | | | | |
| | 5 | 0.25 | | | | | | | |
| | 4 | 0.25 | | | | | | | |
| | 3 | 0.25 | | | | | | | |
| | 2 | 0.25 | | | | | | | |
| | 1 | 0.32 | | | | | | | |

The S-G smoothing method replaces each individual channel value with a weighted arithmetic mean from the actual value and a specified number of neighboring values [8]. The channel length remains the same, so data reduction was not executed. The S-G filter smooths signals by a piecewise adjustment of a polynomial function. The advantage of the S-G filter over the standard smooth function is better maintain the height and the width of peaks of the original signal [9, 10]. The peak stress values from measurements, numerical modelling and signals smoothed with S-G filter are presented in Table 2.

Tab. 2. Comparision of stress values obtained with instrumented rock bolts and numerical modeling

| | Max. stress value in rock bolt no. 1 | | | Max. stress value in rock bolt no. 2 | | |
|---|---|---|---|---|---|---|
| | Recorded stress [MPa] | FEM stress [MPa] | S-G stress [MPa] | Recorded stress [MPa] | FEM stress [MPa] | S-G stress [MPa] |
| Level 5 | 1.33 | 5.30 | 2.06 | 1.45 | 7.73 | 2.86 |
| Level 4 | 1.43 | 5.33 | 2.07 | 1.83 | 7.27 | 2.91 |
| Level 3 | 1.81 | 5.46 | 2.17 | 2.98 | 7.02 | 3.21 |
| Level 2 | 5.94 | 5.60 | 2.31 | 5.67 | 6.87 | 3.36 |
| Level 1 | 1.57 | 5.83 | 2.43 | 2.63 | 6.81 | 3.72 |

From the above analysis one may conclude, that peak stresses obtained from dynamic FEM calculations are nearly coherent with maximum recorded stress changes along bolt's rod. The difference between peak values was 0.34 MPa in bolt no. 1 and 1.20 MPa in bolt no. 2. However, as it was expected, the measured stresses do not fit clearly with stress values obtained from numerical simulations, what is mainly related with low frequency sampling of measuring system. Application of S-G filter (50 points interval) allowed to equalize the resolutions of modeled and measured signals. In result, the modeled stress values were adjusted to measurements. Thus, the average difference between calculated and recorded in situ stresses was 0.20 MPa for rock bolt no. 1 and 0.3 MPa for rock bolt no. 2. Recorded and simulated axial stresses are presented in Figure 5 and Figure 6.

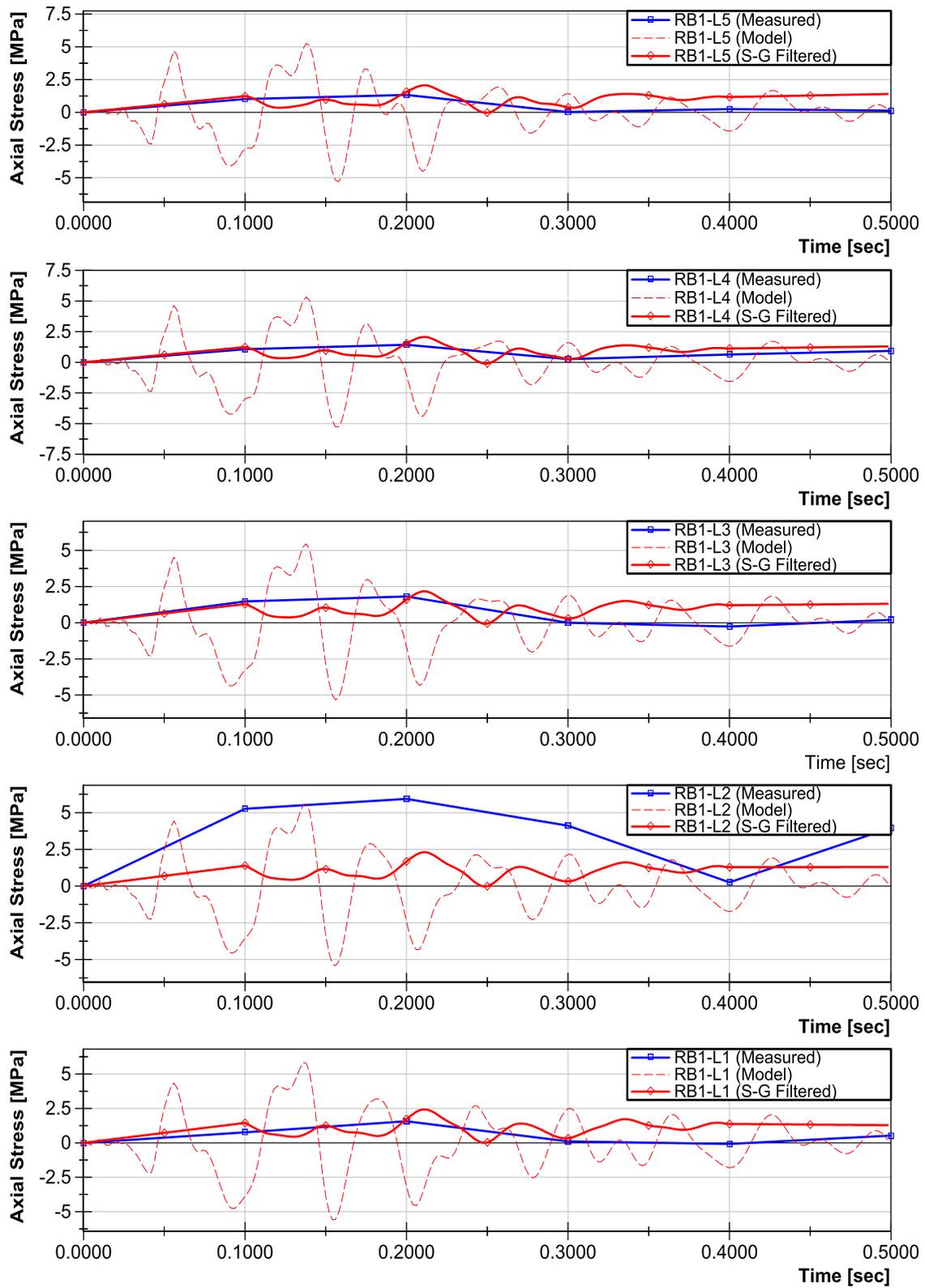

**Fig. 5. Stress values for each measuring level of instrumented rock bolt no. 1**

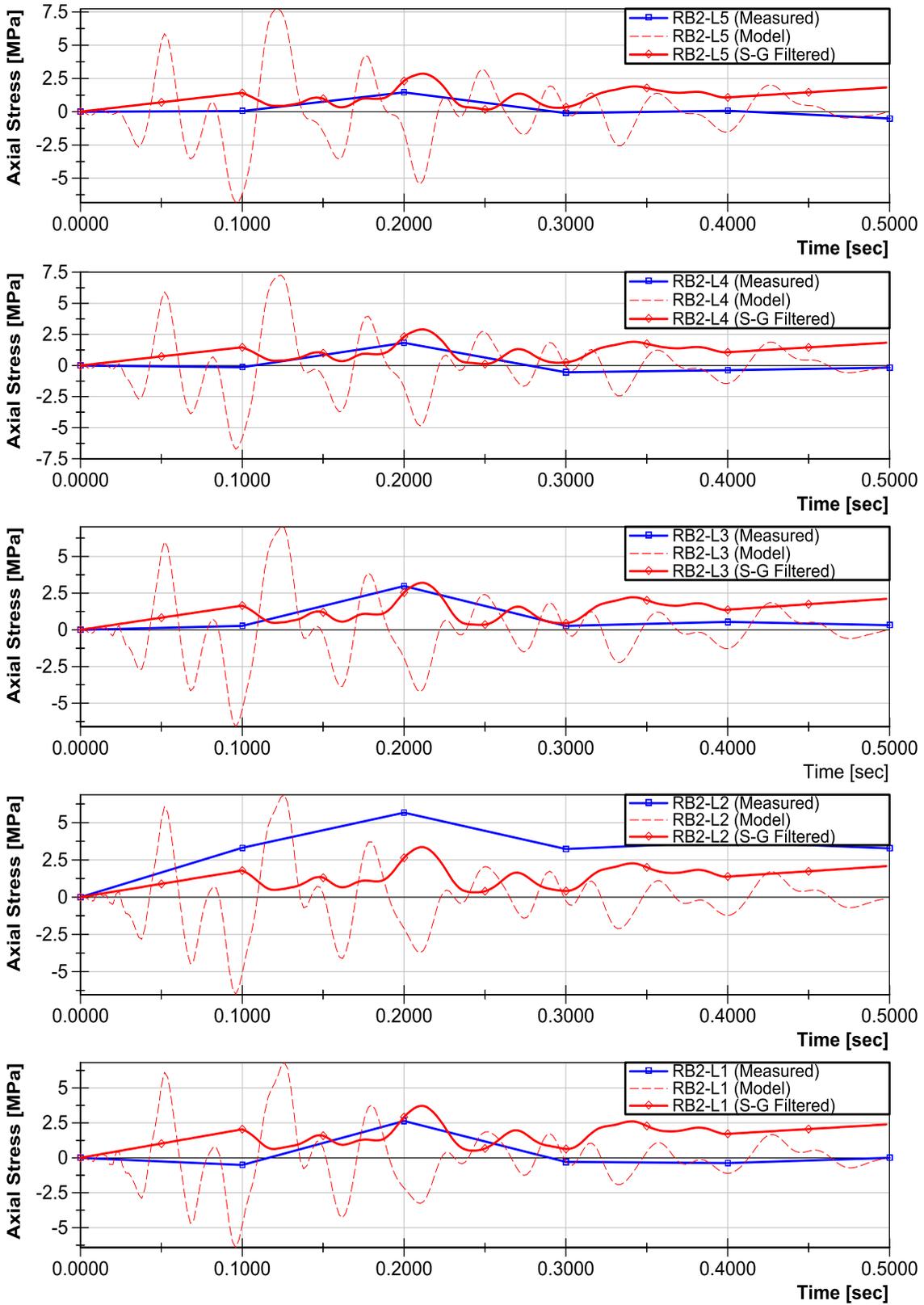

**Fig. 6. Stress values for each measuring level of instrumented rock bolt no. 2**

## 4. CONCLUSIONS

The primary objective of this paper was to assess whether the multi-face production blasting can have negative impact on the roof fall hazard. The influence of blasting works on roof strata condition were considered based on the results of computer simulations and continuous underground stress measurements.

In result, it was found that simultaneous detonation of explosives in a group of mining faces located in close vicinity of paraseismic source, will not significantly affect the fall hazard of continuous rock of the immediate roof strata. Observed axial stress changes did not exceed 8 MPa, despite the small distance (~50 m) between fired mining faces an measuring points. Moreover it was found, that numerical methods could be applied for assessment of stress distribution within the immediate roof strata triggered by blasting works. However, to achieve more precise correlation of results, further underground measurements with higher sampling frequency should be carried out.

## ACKNOWLEDGEMENTS

**This paper has been prepared through the Horizon 2020 EU funded project on "Sustainable Intelligent Mining Systems (SIMS)", Grant Agreement No. 730302.**